\documentclass[preprint,12pt,review]{elsarticle}

\usepackage{amssymb}
\usepackage{amsmath}
\usepackage{color}
\usepackage{spalign}
\usepackage{amsthm}
\usepackage{hyperref}
\usepackage{tabularx}
\usepackage{diagbox}
\usepackage[T1]{fontenc}

\journal{Polymer}

\begin{document}

\begin{frontmatter}



	\title{Molecular dynamics simulations of active entangled polymers reptating through a passive mesh}


	\author[inst1]{Andr\'{e}s R. Tejedor}
	\author[inst1]{Raquel Carracedo}
	\author[inst1]{Jorge Ram\'{i}rez}

	\affiliation[inst1]{organization={Department of Chemical Engineering, Universidad Polit\'{e}cnica de Madrid},
		addressline={Jose Gutierrez Abascal 2},
		city={Madrid},
		postcode={28006},
		state={},
		country={Spain}}




	\begin{abstract}
	
	In this work, we explore the dynamics of active entangled chains using molecular dynamics simulations of a modified Kremer-Grest model. The active chains are diluted in a mesh of very long passive linear chains, to avoid constraint release effects, and an active force is applied to the monomers in a way that it imparts a constant polar drift velocity along the primitive path. The simulation results show that, over a wide range of activity values, the conformational properties of the chains and the tubes are not affected, but the dynamics of the chains are severely modified. Despite not having an explicit tube, the simulations verify all the predictions of the theory about all possible observables very accurately, including a diffusion coefficient that becomes independent of the molecular weight at moderate values of the activity. Overall, this work provides novel information on the study of active entangled polymers, giving a route map for studying this phenomenon and an efficient way of obtaining a polar activity that reproduces the physics of the active reptation theory.
	\end{abstract}

	\begin{graphicalabstract}
		\includegraphics[width=\columnwidth]{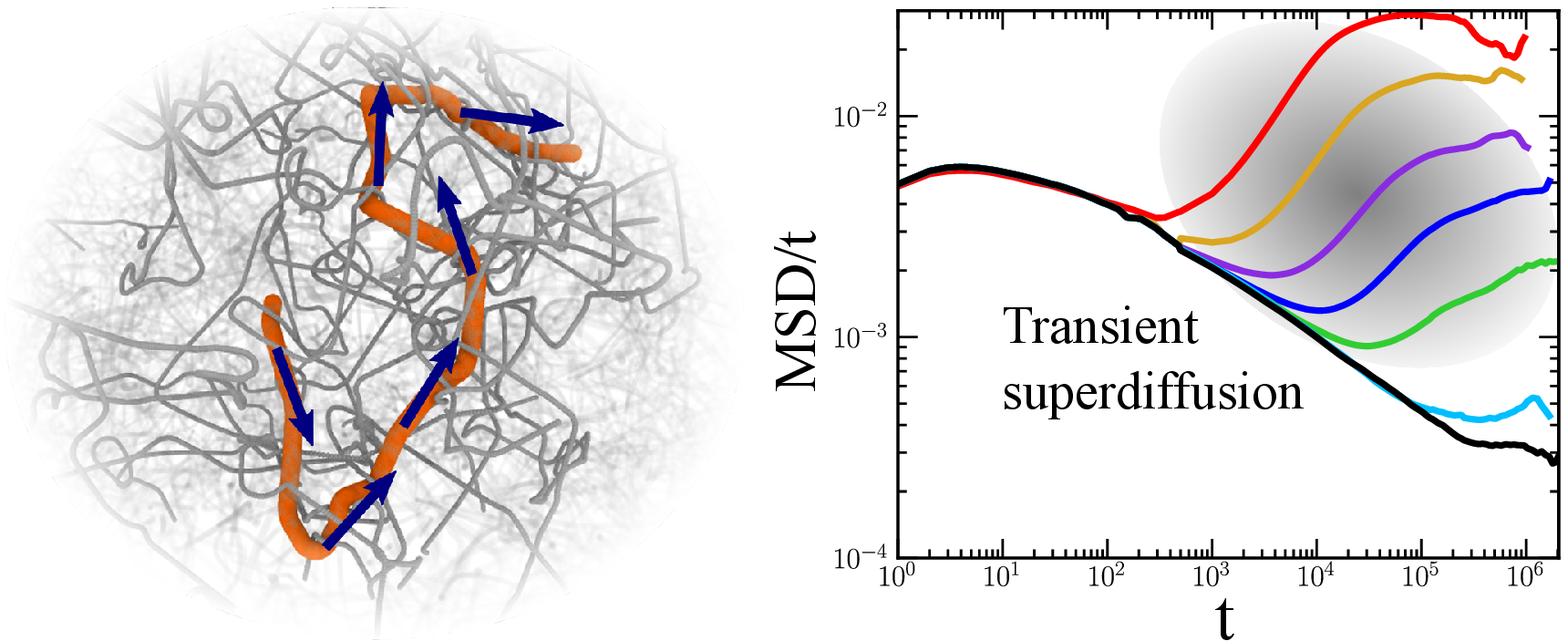}
	\end{graphicalabstract}

	\begin{highlights}
		\item Molecular dynamics of entangled polymers with polar activity through a passive mesh 
		\item Verification of theory for entangled active polymers 
		\item Static properties are not affected in a wide range of activity values
		\item Dynamic modes of relaxation are strongly affected by the activity
		\item Enhanced transport properties of active polymers through a passive mesh
	\end{highlights}

	\begin{keyword}
		Active matter \sep Entangled polymers \sep Active polymers
	\end{keyword}

\end{frontmatter}


\section{Introduction}
\label{sec:introduction}

Active matter is a field of physics that studies out-of-equilibrium systems made of entities that can generate motion from non-thermal energy~\cite{Bechinger_2016, gompper20202020,doostmohammadi2018active}, resulting in a wide variety of collective behavior typically not observed in equilibrium systems~\cite{vliegenthart2020filamentous,roca2022clustering,isele2015,Zottl_2016,negi2022emergent,hagan2016emergent,bar2020self}. Examples of active matter can be found in living systems, such as flocks of birds or colonies of bacteria~\cite{Vicsek_2012}, and also in synthetic systems, such as Janus particles or colloids whose activity can be controlled chemically~\cite{Samin_2015} or by means of electromagnetic fields\cite{Xu_2020}. In particular, active polymers play a crucial role in many functions of the cell, such as transport \cite{hirokawa2009kinesin,2004kinesin}, organization \cite{ganai2014chromosome}, and motility \cite{liu2011force,phillips2012physical}. Several examples of active motion can be mentioned, such as molecular motors (kinesin~\cite{1996kinesin,2004kinesin}, miosyn~\cite{2002miosyn,miosyn2007}), or active filaments (cillia and flagella~\cite{liu2011force,vutukuri2017rational}). In that sense, active polymeric and filamentous structures are fundamental in biological systems, either propelled by molecular motors \cite{vliegenthart2020filamentous,doostmohammadi2018active,keber2014topology,sanchez2012spontaneous,guillamat2017taming} or by themselves \cite{isele2015,Bianco_2018,winkler2017,winkler2020physics}. 
In addition, active systems commonly exhibit enhanced diffusion, which helps to improve many biological processes in which active polymers are involved \cite{phillips2012physical,demirel2010nonequilibrium}. 

The study of active polymers has gained interest in the last decade \cite{isele2015,Bianco_2018,duman2018collective,anand2020conformation,Ghosh2014,Mousavi2019,Kaiser2015,Harder2014a,anand2018structure,anand2019beating}, reaching a relevant position in the more general field of active matter \cite{winkler2020physics,gompper20202020}.  Most of the works to date have focused on active linear polymers in diluted conditions by both theory \cite{philipps2022tangentially,winkler2020physics,martin2018active} and simulations \cite{isele2015,anand2018structure,anand2019beating,paul2022effects,jain2022collapse,Bianco_2018}. However, more complex systems, such as rings \cite{philipps2022dynamics,Mousavi2019}, stars \cite{chuphal2022conformations}, or entangled polymers \cite{Tejedor_2020,Tejedor2019}, are gaining interest. In particular, the study of active entangled polymers may be of great relevance not only in biological systems, but also in industrial processes, because most polymeric materials are processed in melt conditions or in solutions \cite{lim2008processing}. In that sense, as a result of the enhanced transport properties produced by the activity, the viscosity of polymeric fluids can be significantly reduced, as recently demonstrated by experiments \cite{deblais2020rheology} and theory \cite{Tejedor_2020,Tejedor2019}. 

When an isotropic activity acts on the monomers of a polymer chain, it has the effect of a higher effective temperature \cite{Ghosh2014}, which leads to faster diffusion of diluted chains. In entangled polymer chains, an isotropic activity is expected to result in a faster reptation, with no fundamental changes in the dynamical behaviour predicted by the tube theory\cite{doi1988}. However, when a polar activity acts along the contour of the primitive path, i.e., it always points in the direction of one of the ends, identified as head, fundamental changes in the dynamic modes of motion are expected \cite{Tejedor_2020}. As shown by the theory, for high molecular weights, not very large values of the active force are needed to completely change the physics of the chain motion inside the tube. However, the theory makes a series of assumptions similar to those made by the original tube theory\cite{doiI}, i.e., infinitely thin tubes of constant length, no constraint release (CR), no contour length fluctuations(CLF), and the ends can explore all possible orientations freely (isotropic tube segments). Some of these assumptions can be relaxed by running Brownian dynamics simulations of the tube \cite{Wang_2008a, Tejedor_2020} or slip-link models \cite{likhtman2005}. However, to better explore multi-body effects, such as CR, and understand the effect that tubes of finite width have on the dynamics of reptating chains, it is necessary to switch to multi-chain simulations. 

Molecular Dynamics (MD) simulations can be an excellent approach to studying active polymers \cite{gompper20202020}. However, MD simulations of entangled polymers present a high computational cost due to the large molecular weights needed to study well-entangled materials. To reduce that cost, coarse-grained models must be used. In that sense, the Kremer-Grest model \cite{Kremer_1990}, designed to optimize CPU cost while preventing chain crossing, has been successfully used to capture the main predictions of the tube theory \cite{de_Gennes_1971,doiI,doiII,doiIII}. The model has been widely exploited to study entangled polymers at equilibrium \cite{Kremer_1990,zhou2005,likhtman_2007,harmandaris2009}, but has not yet been used to explore the dynamics of active chains under entanglement constraints. 
Very few theoretical or simulation works deal with the dynamics of active entangled chains \cite{Tejedor_2020,Tejedor2019}. Other attempts have focused on two-dimensional systems \cite{vliegenthart2020filamentous} or active particles through a fixed polymer network \cite{kim2022active}. This study is aimed at filling this gap and providing new insights into active entangled polymers through MD simulations.

In this work, we follow our recent study on reptation theory with drift \cite{Tejedor_2020,Tejedor2019}, to verify the main predictions stated there by MD simulations. In the theoretical work, the activity is introduced to act along the primitive path in the directions of one of the chain ends. Here, we carry out simulations of the Kremer-Grest model, implementing the activity in a way that mimics the theory. To discard the effect of constraint release, which is not considered in the theory, we simulate self-propelled, diluted, and mildly entangled chains moving through a mesh made of long passive polymers. We first review the main hypothesis and results of the theory (Sec. \ref{subsec: theory}), providing an estimation of the P{\'e}clet number, which was absent in the original theory (Sec. \ref{subsec: Peclet}), and give details about the simulation model used to reproduce the underlying physics (Sec. \ref{subsec: KG}). The basic hypotheses of the theory are confirmed by analyzing the conformation of the active polymers and the structure of the tube (Sec. \ref{subsec: conformation}). The results related to dynamical properties are presented following our original work \cite{Tejedor_2020}, calculating the mean square displacement of the center of mass (Sec. \ref{subsec: MSD of COM}), the monomeric mean square displacement (Sec. \ref{subsec: MonomericMSD}), the end-to-end relaxation (Sec. \ref{subsec: EteRelax}), the tangent-tangent relaxation and tube segment survival function (Sec. \ref{subsec: TanTan}), and finally the dynamic structure factor (Sec. \ref{subsec: DSF}). Overall, we aim to explore the dynamical response of entangled polymers subjected to a polar active force by means of Molecular Dynamics simulations and verify the most important predictions of the previously published theory.

\section{\label{sec: methods}Materials and methods.}

\subsection{Reptation of active entangled polymers}
\label{subsec: theory}

In this section, we provide a brief overview of the analytical theory of active entangled polymers that we have recently published \cite{Tejedor2019,Tejedor_2020}. We extend the original model of Doi and Edwards \cite{doi1988} to the case where an active force is applied to the monomers so that it imparts a constant polar drift velocity $c$ to the primitive chain, pointing from tail to head along the primitive path. In the original work\cite{Tejedor2019}, we hypothesize about possible ways to enforce this activity. Constraint release is not considered, so the model represents the motion of an active entangled chain moving through a mesh of fixed obstacles or through a mesh of passive chains of much higher molecular weight. To ensure the basic assumptions of the tube model, we limit the drift value to $c\ll a/\tau_e$, where $a$ is the diameter of the tube and $\tau_e$ is the Rouse time of an entanglement. In this way, we can safely assume that the end segments of the tube are free to explore all possible orientations, and thus that all tube segments are renewed as soon as they are visited by either of the chain ends. As a consequence of the isotropic renewal of the tube segments, the static properties of the chain and the tube (end-to-end vector, radius of gyration, distance between monomers, etc.) are not affected by the drift.

By introducing a constant drift $c$ along the primitive path, the characteristic time for a chain to escape the tube goes from $\tau_d \propto N^3$ (for pure reptation) to $\tau_d \propto N$ (for a relevant drift), where $N$ is the number of monomers or the molecular weight of the polymer. The drift $c$ does not need to be excessively large in order to be relevant, since the critical value of $c$ to observe drift-governed dynamics scales as $c \propto N^{-2}$. Thus, no matter how small the value of the drift is, there will always be a sufficiently long linear chain whose dynamics will be dominated by the activity instead of the diffusive reptation motion. This change in the scaling of the terminal time immediately affects all other dynamical observables of the system, such as viscosity ($\eta\propto N$), end-to-end relaxation (also proportional to $N$), and self-diffusion coefficient (independent of the molecular weight), as can be shown by using a simple scaling argument. We know that every time the chain escapes the tube $\tau_d \propto N$, the center of mass must have moved a mean square distance which is proportional to the coil size $\langle R^2 \rangle \propto N$. As a result, the diffusion coefficient, estimated as $D_G \propto \langle R^2\rangle / \tau_d$ is  independent of the molecular weight, whereas it decreases as $D_G\propto N^{-2}$ in pure reptation. 

Some other relevant results of the theory that we intend to check by means of Molecular Dynamics simulations are as follows: (i) The polar character of the drift, directed from tail to head, yields a tube survival function that is asymmetrical with respect to the center of the tube. This is to be expected, since the tube segments closer to the tail are more rapidly renewed than those closer to the head, because the tail end, driven by the drift, has a higher tendency to move into the primitive path than the head. (ii) Interestingly, despite this asymmetry in the tube survival function, the mean square displacement of the segments along the chain $g_1(s,t)$ is fully symmetric with respect to the center of the tube. The slowest monomer is always the middle monomer, as in pure reptation, and the motion of monomers is subdiffusive at all times before reaching the terminal Fickian regime. (iii) The famous reptation power laws in the mean square displacement of the middle monomer, predicted by Doi and Edwards \cite{doi1988} and verified many times in Molecular Dynamics simulations \cite{Kremer_1990, P_tz_2000}, progressively disappear as the drift increases. For high drift values, $g_1(s,t)$ transitions directly from the Rouse slope $g_1 \propto t^{0.5}$ at $\tau_e\ll t\ll \tau_R$ to the Fickian slope $g_1 \propto t$, without showing any signs of the characteristic reptation slope $g_1 \propto t^{0.25}$ between $\tau_R$ and $\tau_d$. This occurs when the terminal time due to the drift $\tau_d = Za/c$ is smaller than the Rouse time $\tau_R = Z^2 \tau_e$, \emph{i.e.} when $c >  a/Z\tau_e$. (iv) Finally, although the segmental motion is subdiffusive, the mean square displacement of the center of mass shows a transient superdiffusive scaling before reaching the therminal Fickian regime at $\tau_d$. 

Overall, we can consider the activity as a force with a constant module and direction tangent to the primitive path that acts on the polymer segments. As such, the sum of all active forces is a global force that acts on the center of mass and points in the direction of the end-to-end vector. Thus, the dynamics of the center of mass are expected to be very similar to those of an active Brownian particle (ABP) \cite{Howse_2007}. In ABPs, a constant force acts on a colloid in a direction whose orientation changes by rotational diffusion. Here, the direction of the global force also changes by orientational diffusion, and the characteristic reorientation time is given by the disengagement time $\tau_d$ of the polymer chain. However, the force is not constant, as the end-to-end vector is a fluctuating 3D Gaussian variable.

\subsection{Estimation of the Peclet number}\label{subsec: Peclet}

When active colloids or molecules are considered, it is important to quantify the strength of the activity with respect to the equilibrium Brownian motion by means of the P\'{e}clet number (Pe). In principle, two different Pe numbers can be defined, one related to the translational motion and the other to the rotational motion of the molecule:
\begin{eqnarray}
		\mathrm{Pe}_t &= V R_g /D_G  \\
		\mathrm{Pe}_r &= V \tau_d /R_g
\end{eqnarray}
where $V$ is the drift velocity imparted by the activity to the center of mass, $R_g$ is the molecular radius of gyration, $D_G$ is the self-diffusion coefficient of passive chains (proportional to $Z^{-2}$ \cite{doi1988}) and $\tau_d$ is the disengagement time, which is equivalent to the terminal time of the end-to-end vector relaxation or the rotational relaxation time. We can estimate the value of both $Pe$ numbers using the results of the theory. As explained above, in the theory we assume that the range of activities considered is small, and thus the conformational properties of the chains are not affected. Assuming Gaussian statistics, the chain size is $R_g^2=Za^2/6$, where $Z$ is the number of entanglements and $a$ is the diameter of the tube. For passive chains, the terminal time is given by the disengagement time $\tau_d$ from reptation theory. However, when the activity imposes a constant drift velocity $c$ along the primitive path, the terminal time changes to $\tau_{d}=Za/c$ (the time it takes for the tail monomer to move over the contour length of the primitive path with velocity $c$). 

In order to estimate the value of $V$, we can consider the motion of the center of mass of the polymer chain as if it were an active Brownian particle (ABP), as discussed above. The diffusion coefficient of an ABP is \cite{Zottl_2016}
\begin{equation}
	D_{ABP} = D_G + \frac{V^2\tau_d}{3} \approx \frac{V^2Za}{3c} = \frac{ca}{6}
\end{equation}
where we have considered that $D_G$ is negligible when the activity is large and, in the last equality, we have introduced the asymptotic value of the diffusion coefficient for an active entangled chain \cite{Tejedor_2020}. The resulting active velocity of the center of mass is
\begin{equation}
	V \propto \frac{c}{\sqrt{Z}}
\end{equation}
Therefore, the translational $\mathrm{Pe}$ is:
\begin{equation}
	\mathrm{Pe}_t\propto  cZ^2
\end{equation}
and the rotational $\mathrm{Pe}$ is:
\begin{equation}
	\mathrm{Pe}_r= \sqrt{6}
\end{equation}
It is interesting to note that the translational P\'{e}clet number $\mathrm{Pe}_t$ depends on the square of the molecular weight. This is consistent with the results of our previous work, where we found that the expression $cZ^2$ governs whether the drift dominates over diffusion \cite{Tejedor2019,Tejedor_2020}. In contrast, the rotational $\mathrm{Pe}$ number is constant and does not depend on the molecular weight. 

\subsection{Coarse-grained model and simulations}\label{subsec: KG}

We use the Kremer-Grest model \cite{Kremer_1990} to simulate active entangled polymers. Each polymer chain is described as a linear arrangement of $N$ repulsive Lennard-Jones beads \cite{Weeks1971} connected by finitely extensible nonlinear elastic (FENE) bonds \cite{bird1987dynamics}. The Weeks-Chandler-Andersen (WCA) potential or repulsive Lennard-Jones is given by:
\begin{equation}
    U_{\text{WCA}}(r)=\spalignsys{
        4\epsilon\left[\left(\frac{\sigma}{r}\right)^{12}-\left(\frac{\sigma}{r}\right)^{6}\right]+\epsilon,\qquad r<2^{1/6}\sigma;
        0,  \qquad\text{otherwise,}
   }
\end{equation}
where $r$ is the distance between the beads, $\epsilon$ is the Lennard-Jones energy, and $\sigma$ stands for the bead size. The bonded FENE potential has the form
\begin{equation}
    U_{\text{FENE}}(r)=-k\frac{R_0^2}{2}\log\left[1-\left(\frac{r}{R_0}\right)^2\right],
\end{equation}
where $R_0=1.5\sigma$ is the maximum extension value of the spring and $k=30.0\epsilon/\sigma^2$ is the spring constant. For the simulation of active polymers, the FENE bonded potential is rather stiff, and it may not be suitable for extremely high values of $\mathrm{Pe}$ (in fact, most works working with simulations of active polymers use harmonic bonds \cite{Bianco_2018, winkler2020physics}). However, we want to enforce chain uncrossability, for which the Kremer-Grest model is known to work well. In addition, we are interested in a range of activities that are significant in affecting the reptation dynamics but still small to maintain the isotropic orientation of tube segments along the contour of the polymer, and prevent any chain stretching. Thus, we expect that the tube parameters and the bond lengths will be barely affected by the activity. For simplicity, in the following, we use reduced Lennard-Jones units so that $\epsilon=1$, $m=1$, $\sigma=1$, and $\tau=\sqrt{\sigma^2m/\epsilon}=1$ give the units of energy, mass, distance, and time, respectively. We run simulations in the NVT ensemble using a Langevin thermostat and keeping the temperature at $T=k_B/\epsilon$, where $k_B$ is the Boltzmann constant. All simulations are run using the LAMMPS software \cite{LAMMPS}. 

In the theory \cite{Tejedor2019, Tejedor_2020}, we hypothesize a drift that acts tangentially to the primitive path of the virtual tube. Now, we aim to mimic the same behavior, but in multi-chain MD simulations the tube is not well defined to set the activity accordingly. A primitive path could be constructed every several time steps of the MD simulation using any of the well-established algorithms \cite{sukumaran2005, Kroger_2005}, but this would have a critical impact on the efficiency of the simulations. Instead, we introduce drift as a force proportional to the bond (with a proportionality constant $f_c$) acting on each bead and pointing in the direction of the next monomer toward the `head' end, so each bead has the following contribution from the active force:
\begin{equation}\label{eq: activeforce}
    f_i=f_c\cdot(r_{i+1}-r_{i-1}).
\end{equation}
The net active force acting on the chain is parallel to the end-to-end vector, in agreement with the total active force used in the theory. In the strands between entanglements, we expect that the component of the active force perpendicular to the primitive path will average to zero, so that the resulting force on the monomers in that entanglement will have the direction of the primitive path. 

\begin{figure*}[hbt!]
	\centering
	\includegraphics[width=\columnwidth]{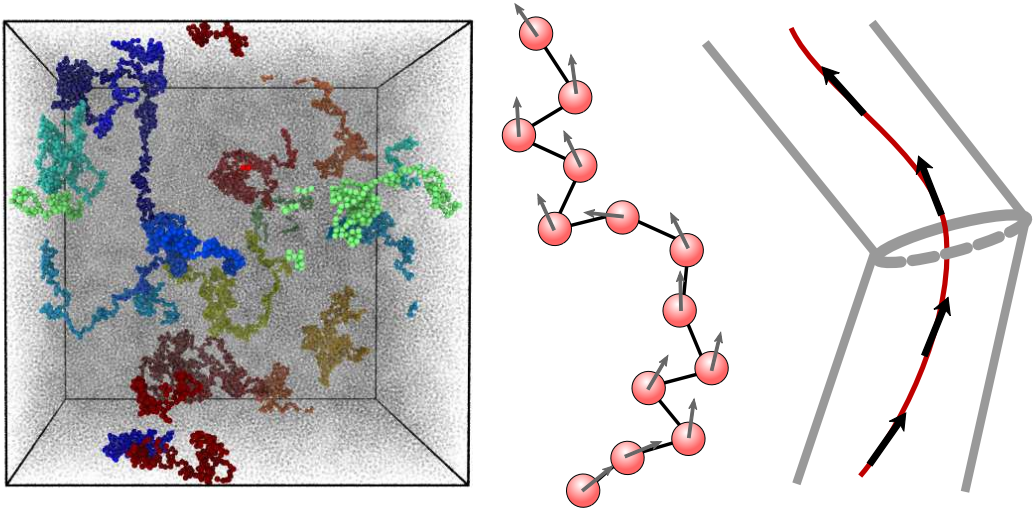}
	\caption{Snapshot from simulations showing the tracer active chains in different colors and the passive chains of the mesh in transparent gray color. 
	Sketch of the force in our coarse-grained model and effective force in the tube representation where the force acts tangentially to the primitive path.}
	\label{fig: Sketch}
\end{figure*}

In the analytical theory \cite{Tejedor2019, Tejedor_2020} we consider an active chain that diffuses by reptation through a mesh of fixed obstacles and thus neglects the effects of contour length fluctuations and constraint release. To mimic those conditions, we embed an ensemble of $N_c$ active chains of length $N$ diluted inside a mesh of $N_{c,m}$ passive chains of much higher molecular weight ($N_m=1000$), so that the active chains do not see each other. The terminal time of the passive long chains is much longer than that of the active ones, so, effectively, the entanglement network felt by the active chains can be considered to remain fixed during the relaxation of active polymers (we provide all the system parameters in Table \ref{tab: SystemParameters}). In the MD simulations, it is not possible to turn off contour length fluctuations, and the tube has a certain finite width, so this work tests whether the theory is still valid in the presence of CLF and fat tubes. The agreement of the theory with CLF was partly tested in a previous work \cite{Tejedor_2020}, where the results of the theory were compared with Brownian dynamics simulations of single active entangled chains. In this work, we extend the comparison to the more realistic scenario of a multi-chain simulation.

\begin{table}[htb!]
    \centering
    \begin{tabular}{|c|c|c|c|c|}
        \hline
         $\quad N_c\quad $ & $\quad N\quad $ & $\quad N_m\quad $ & $\quad N_{c,m}\quad $ & $\quad L/\sigma\quad $ \\
         \hline
         \hline
        70 & 50 & 1000 & 123 & 53.0 \\
        \hline
        25 & 100 & 1000 & 127 & 53.4\\
         \hline
        10 & 200 & 1000 & 145 & 55.7\\
        \hline
        10 & 300 & 1000 & 268 & 68.3\\
        \hline
        10 & 400 & 1000 & 414 & 79.0\\
        \hline
        10 & 500 & 1000 & 579 & 88.3\\
        \hline
    \end{tabular}
    \caption{System parameters used in this study, including the number of beads per active chain $N$, the number of active probe chains $N_c$, the number of beads per passive chain $N_m$, the number of passive mesh chains $N_{c,m}$, and the length of the simulation box $L$.}
    \label{tab: SystemParameters}
\end{table}
We randomly placed all the chains inside a cubic simulation box, with an overall monomeric density of $\rho=0.85$. For each molecular weight $N$, the size of the box is determined so that the active chains are in the dilute regime (the mean distance between the centers of mass of the active chains is much larger than their mean size). The system is equilibrated using the fast push-off method \cite{Auhl_2003,Sliozberg_2012}, without introducing any activity. After the systems are equilibrated, the active force is introduced in the short probe chains only, and the systems are run until the mean square displacement of the center of mass of the active chains reaches the terminal Fickian regime. 

\subsection{Calculation of observables}

Care has been taken to ensure that the activity values used in this work do not produce any phase separation or strong density fluctuations in the system. For that purpose, we have monitored the density and radial distribution functions of our simulations of active systems and checked that they are identical to those of their passive counterparts.

All the dynamical observables studied (center of mass and segmental mean square displacement, end-to-end vector relaxation, tube survival function, and dynamic structure factor) have been calculated using a multiple tau correlator approach, which allows the calculation of time correlation functions on the fly during a simulation without a strong impact on the efficiency. \cite{Ramirez_2010}.

The entanglement network and the primitive path statistics have been calculated using the primitive path analysis algorithm, implemented in a modified version of LAMMPS~\cite{everaers2004,hagita2021,sukumaran2005} (see Section SIB in the Supplementary Material for further details).

\section{Results and discussion}\label{sec: Results}

\subsection{Effect of activity on the structure and conformation}\label{subsec: conformation}

Although the levels of activity studied in this work are sufficient to greatly affect the dynamical properties, in this section we show that the selected values of $f_c$ are not high enough to affect the static properties of the chains and the tube. 

According to the theory \cite{Tejedor_2020}, if the drift velocity $c$ is much smaller than $a/\tau_e$, then the end segments are free to explore all possible orientations and the tube remains a random walk at all times. The Rouse time and the number of monomers in one entanglement in the Kremer-Grest model can be estimated as $\tau_e = 5800$ and $N_e=72$, respectively \cite{likhtman_2007}. With a characteristic ratio $C_\infty=1.88$ \cite{Sliozberg_2012} and an average bond length $b=0.97$ \cite{Kremer_1990}, we can estimate the radius of the tube as $a=\sqrt{C_\infty N_e}b=11.3$. Thus, the limiting drift is given by $a/\tau_e=1.95\cdot10^{-3}$, which is much larger than the drift imparted by the maximum $f_c$ used in this work, as we prove later after computing the self-diffusion coefficient $D_G$ and the drift velocity $c$ with Eq. \eqref{eq: DG} (see Table \ref{tab: MapFc}). Therefore, we expect that the tube remains a random walk at all times, and the static properties of the chains and the tube should be the same as in the passive case. 

\begin{figure}[hbt!]
	\centering
	\includegraphics[width=\columnwidth]{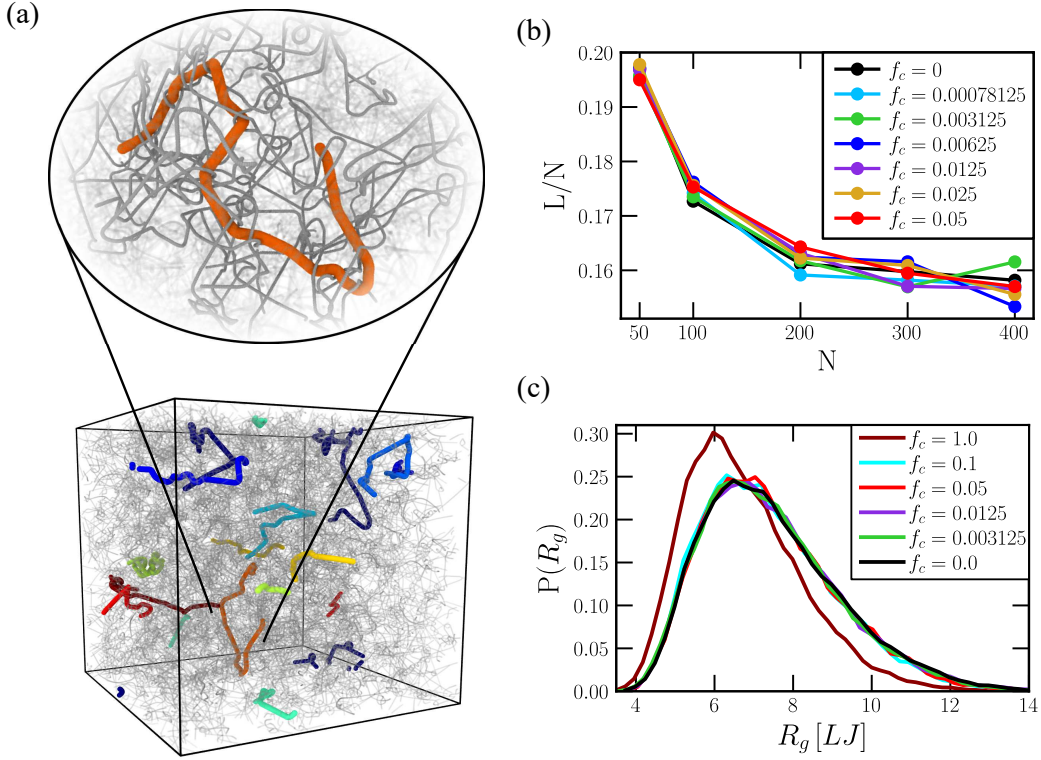}
	\caption{(a) Snapshots from the primitive path analysis for a system with probe chains of $N=400$ beads and zero activity. Long chains forming the mesh are blurred to allow visualization of the coloured active chains. The zoom snapshot highlights the orange chain as well as representative long chains that form the corresponding virtual tube. The bead sizes of the mesh chains have been reduced for simplicity. (b) Contour length of the tube divided by number of beads as a function of the polymer length ($N$) for every studied activity. (c) Probability distribution of the radius of gyration of active chains of $N=200$ at activities ranging from $f_c=0$ to 1.}
	\label{fig: PPAandRg}
\end{figure}

In Fig. \ref{fig: PPAandRg}c) we show the probability distributions of the radii of gyration of the active chains for $N=200$ at different activities, showing that the coil size is not affected by the active force. The $R_g$ distributions for the active chains remain equal to those of the passive case, within the noise of the results due to the small number of active chains in the system. Only for very large activities ($f_c\gtrsim1$) the chains do start to collapse inside their tubes, and the distribution of $R_g$ is shifted to smaller values. Bearing this in mind, we restrict all our simulations to active forces below $0.1$ to ensure that the static properties are not affected by the active force. We have also checked that the activity, although it has a marked polar effect, does not introduce any noticeable differences in the conformations of the head and tail halves of the chains for the activities we are considering (data not shown). 

In addition to the dimensions of the polymer coil, it is important to verify that the characteristics of the tube, such as the tube diameter and tube length, are not affected by the activity. The topological tube is constructed by passive chains with larger molecular weight than the probes, so the mesh structure is expected to remain constant during the disengagement time of the active chains. Thus, the tube diameter, which measures a property of the mesh, should also remain constant and unaffected by the activity. To test whether the length of the tube changes with the activity, we have performed a primitive path analysis (PPA) \cite{everaers2004, sukumaran2005} of all simulated systems to extract the primitive path length $L$ of the active chains, and calculated the tube diameter as $\langle R^2\rangle/L$, where $\langle R^2\rangle$ is the mean squared end-to-end distance of the active chains. A sample snapshot of a simulation box after the PPA has been performed is depicted in Fig. \ref{fig: PPAandRg}a), showing the primitive paths of the active chains (in color) and the entanglements created by the passive long chains (in grey, with a detailed example shown in the zoom). The average length of the primitive path $L$ of active chains divided by the molecular weight $N$ (see Fig. \ref{fig: PPAandRg}) demonstrates that the tube structure is not affected by the activity, as expected.

\subsection{Diffusion and segmental motion}

\subsubsection{Mean square displacement of the center of mass and diffusion coefficient}\label{subsec: MSD of COM}

\begin{figure}[hbt!]
	\centering
	\includegraphics[width=\columnwidth]{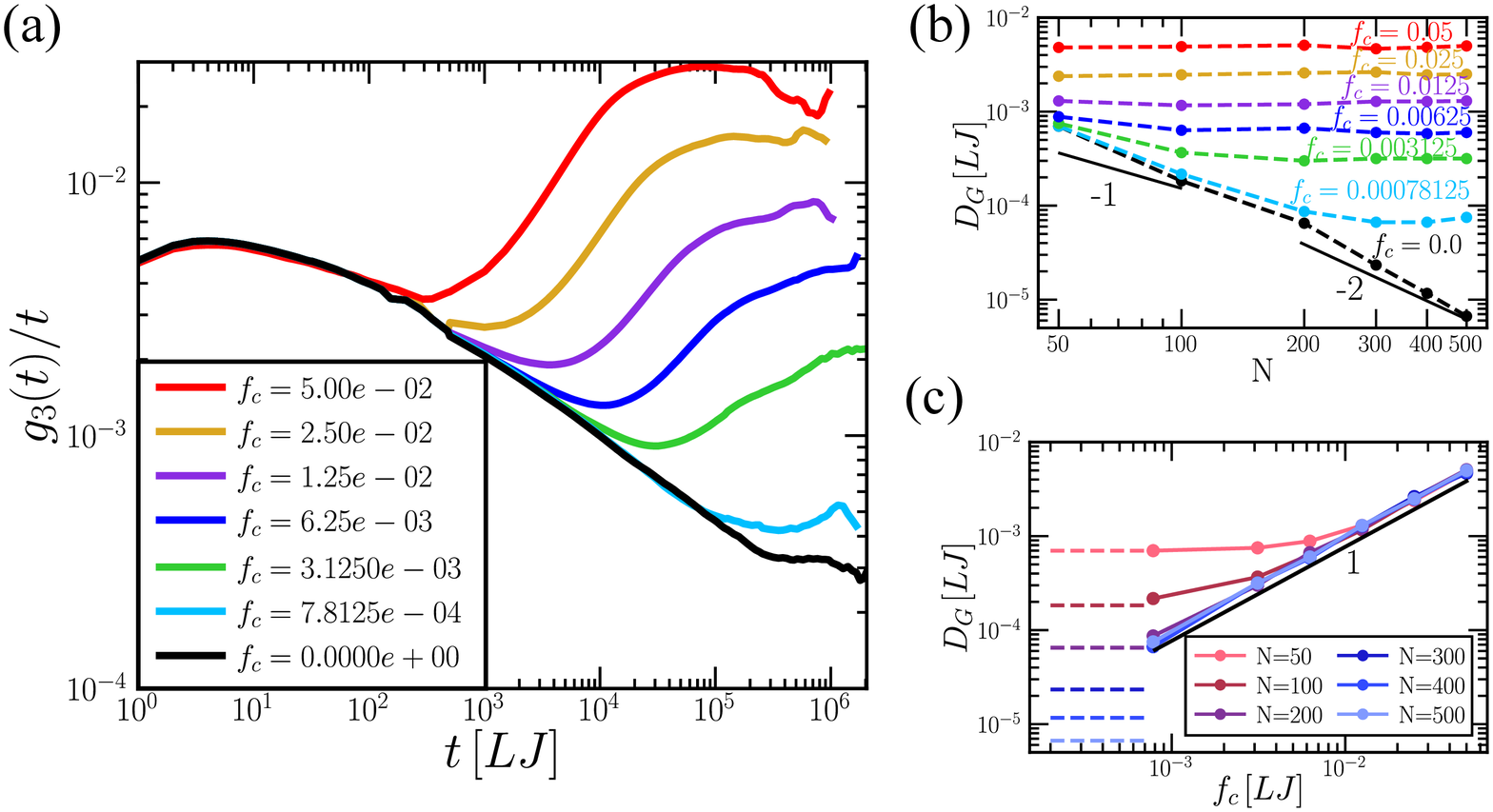}
	\caption{(a) Mean square displacement of the center of mass divided by time of active chains with $N=200$ at all activities studied in this work. (b) Diffusion coefficient of active probe chains measured from $g_3(t)$ for all the activities studied vs chain length. Dashed lines are included as a guide for the eye. Solid black lines indicate the characteristic power laws from the Rouse (-1) and tube theory (-2). (c)  Diffusion coefficient of the active probe chains for all polymer lengths studied vs. the introduced active force in Lennard-Jones units. Solid lines serve as a guide for the eye. Horizontal dashed lines indicate the diffusion coefficient for the passive case corresponding for each chain length.}
	\label{fig: FigDiffAndg3}
\end{figure}

The drift-governed reptation motion imparted by the active force is expected to enhance the diffusion of the entangled probe chains. We can study this activity-induced behavior by calculating the mean square displacement of the center of mass $g_3(t)=\big\langle\big(\mathbf{r}_{CM}(t)-\mathbf{r}_{CM}(0)\big)^2\big\rangle$ (see Fig. \ref{fig: FigDiffAndg3}). At a very early time, the mean square displacement (MSD) of the center of mass is dominated by reptation, due to the scaling $\langle s^2\rangle\propto t$ of diffusion and $\langle s^2\rangle\propto t^2$ of a motion dominated by a drift velocity, where $s$ is the distance travelled by the chain along the primitive path. However, according to the theory \cite{Tejedor_2020}, $g_3(t)$ should show transient superdiffusion for about two decades in time before reaching the terminal Fickian regime. This superdiffusive regime should be more visible for longer chains since the action of the drift increases with the square of the molecular weight (see Fig. 8 in \cite{Tejedor_2020}). In Fig. \ref{fig: FigDiffAndg3}, we have divided the MSD of the center of mass by time to highlight the Fickian regime and the superdiffusive transient regimes for active chains with $N=200$ at all activities. As expected, passive chains are subdiffusive and only reach diffusive behavior after the disengagement time. Instead, active chains depart from the passive MSD at increasingly early times as the activity grows, show a superdiffusive regime that is more pronounced for higher activities, and reach a Fickian regime plateau (\emph{i.e.} a self-diffusion coefficient) that is higher as the activity increases, qualitatively supporting the predictions made by the theory, which is remarkable considering that the chains are only mildly entangled for $N=200$ and both the effects of CLF and the finite width of the tube may be very important. 

From $g_3(t)$ we can easily calculate the diffusion coefficient using $D_G=\lim_{t\rightarrow\infty}g_3(t)/6t$ (see, for example, Eq. (6.40) in \cite{doi1988}), which in our case corresponds to the difusive behavior when $g_3(t)/t$ reaches a plateau. The theory predicts that $D_G$ should be independent of the molecular weight and proportional to the drift velocity \cite{Tejedor2019,Tejedor_2020}. In Fig. \ref{fig: FigDiffAndg3}(b), the self-diffusion coefficient is shown as a function of the molecular weight for all activities studied in this work. As can be seen, $D_G$ is imperatively determined by the activity and, while passive chains (black dots) show the characteristic transition from Rouse-like behavior ($D_G\propto N^{-1}$) to the reptation-dominated regime ($D_G\propto N^{-2}$) at a molecular weight of approximately $N\sim 200$ (corresponding to roughly 3 entanglements), the behavior of active entangled chains is radically different. At moderate activities, the diffusion coefficient becomes independent of the molecular weight and is constant for each value of the activity. This result is very interesting not only because it supports the predictions of the theory, but also because there are no signs of a transition between the unentangled and the entangled regimes. Previous studies on unentangled dilute chains \cite{Bianco_2018,winkler2017,isele2015} show that the diffusion coefficient of active polymers with polar activity directed along the contour of the chain is independent of the molecular weight, and it is remarkable that this behavior persists still when active unentangled chains move in a dense medium where hydrodynamic interactions are screened. However, it is not expected that the diffusion coefficient does not change when entanglements arise, even though the dynamics of the system is governed by completely different physics, since unentangled and entangled chains are subjected to completely different geometrical constraints, and the fact that an active force tangentially applied to the chain contour yields exactly the same diffusion coefficient in both cases deserves further investigation.

\begin{table}[htb!]
    \centering
    \begin{tabular}{|c|c|c|c|c|c|c|}
        \hline
         \diagbox[width=27mm]{$\mathbf{f_c\,(\epsilon/\sigma)}$}{$\mathbf{c\,(\sigma/\tau)}$} & \scalebox{0.84}{$N=50$} & \scalebox{0.84}{$N=100$} & \scalebox{0.84}{$N=200$} & \scalebox{0.84}{$N=300$} & \scalebox{0.84}{$N=400$} & \scalebox{0.84}{$N=500$} \\
         \hline
        7.8125e-4 & 3.30e-5 & 1.03e-5 & 4.13e-6 & 3.17e-6 & 3.17e-6 & 3.57e-6 \\
        \hline
        3.125e-3 & 3.57e-5 & 1.75e-5 & 1.43e-5 & 1.51e-5 & 1.51e-5 & 1.51e-5 \\
         \hline
        6.25e-3 & 4.20e-5 & 3.01e-5 & 3.17e-5 & 2.86e-5 & 2.78e-5 & 2.86e-5 \\
        \hline
        1.25e-2 & 6.19e-5 & 5.55e-5 & 5.71e-5 & 6.11e-5 & 6.11e-5 & 6.19e-5 \\
        \hline
        2.5e-2 & 1.13e-4 & 1.17e-4 & 1.23e-4 & 1.25e-4 & 1.17e-4 & 1.19e-4\\
        \hline
        5.0e-2 & 2.29e-4 & 2.34e-4 & 2.42e-4 & 2.22e-4 & 2.30e-4 & 2.38e-4 \\
        \hline
    \end{tabular}
    \caption{Effective drift velocity output for all activities and active polymer lengths, expressed in Lennard-Jones units.}
    \label{tab: MapFc}
\end{table}

When plotted as a function of the applied activity (Fig. \ref{fig: FigDiffAndg3}c), the diffusion coefficient is perfectly proportional, as predicted by the theory, except for short chains and very small activities, when the drift is not large enough to completely dominate over reptation dynamics. Furthermore, the terminal diffusion coefficient is not affected by the activity in some cases for $N=50$ and $N=100$, so the active force is not high enough to improve the transport properties. The qualitative agreement of this results with the predictions of the theory, confirm that the choice of implementation of the activity in the MD simulations is efficient in mimicking the reptation with drift physics of the theory \cite{Tejedor_2020}, despite not having a precise definition of the tube in multi-chain simulations. The force considered in this work ($f_c$, see Eq. \eqref{eq: activeforce}) results in a constant drift ($c\propto f_c$) along the primitive path. Therefore, we can calculate the effective velocity along the contour of the primitive path using Eq. (20) of the theory \cite{Tejedor_2020}
\begin{equation}\label{eq: DG}
    D_G=\frac{c}{6}\coth{\left(\frac{3\pi^2cZ^2}{2}\right)},
\end{equation}
which in the limit of drift governing the dynamics can be simplified as $D_G=c/6$. Using this relation, we can infer the relation between the input $f_c$ and the effective drift $c$ and check the limits of our theory for this simulation model. 
In Table \ref{tab: MapFc} we provide the drift velocities obtained for all systems studied. 
First, it is important to note that none of the drifts in the table exceeds the limiting value of $c=1.95\cdot 10^{-3}$ derived in Sec. \ref{subsec: conformation}.
In addition, as stated above, when the activity governs over reptation, the drift velocity along the tube is directly proportional to the active force, which is a non trivial result. 
We can estimate the threshold value of $f_c$ for which the activity starts to dominate over reptation ($f_{min}$) and 
the upper limit for the theory assumptions to hold ($f_{max}$, see the Supplementary Material for further details on this calculation). Note that the calculation of $f_{max}$ comes from a local argument related to the isotropy of the end tube segments, so it does not depend on the molecular weight. On the other hand, the calculation of $f_{min}$ involves the diffusion coefficient of the whole polymer (Eq. (21) in the original work~\cite{Tejedor_2020}), so that it does depend on the molecular weight of the chain. Both threshold limits for the activity are shown on Table \ref{tab: LimitsFc}. It is clear that all the values of the activity studied in this work are below the upper threshold. However, some of the activities are not relevant for the smallest molecular weights, in agreement with the scalings shown in Fig. \ref{fig: FigDiffAndg3}b) and c). In fact, the $R_g$ distribution for $f_c=1.0$ shown in Fig. \ref{fig: PPAandRg}c) clearly proves that the activity value is out of the limits where the theory is applicable. 
\begin{table}[htb!]
    \centering
    \begin{tabular}{|c|c|c|c|c|c|c|}
        \hline
          & \scalebox{0.84}{$N=50$} & \scalebox{0.84}{$N=100$} & \scalebox{0.84}{$N=200$} & \scalebox{0.84}{$N=300$} & \scalebox{0.84}{$N=400$} & \scalebox{0.84}{$N=500$} \\
         \hline
        $\mathbf{f_{min}\,(\epsilon/\sigma)}$ & 0.015 & 0.0038 & 9.5e-4 & 4.3e-4 & 2.4e-4 & 1.5e-4\\
        \hline
        $\mathbf{f_{max}\,(\epsilon/\sigma)}$ & 0.41 & 0.41 & 0.41 & 0.41 & 0.41 & 0.41 \\
         \hline
    \end{tabular}
    \caption{Upper and lower limiting values of the parameter $f_c$ for all the studied systems.}
    \label{tab: LimitsFc}
\end{table}

\subsubsection{Monomeric MSD}\label{subsec: MonomericMSD}

As already stated in Section \ref{subsec: theory}, both the theory and Brownian dynamics simulations detailed in our previous work \cite{Tejedor2019,Tejedor_2020} predict that: (i) Even though the activity induces a polarity along the chain, the mean square displacement of monomers shows a head-tail symmetry, i.e., the head monomer, which constantly creates new tube segments, and the tail monomer, which follows the generated primitive path from the other end of the chain, have exactly the same form of the monomeric mean square displacement function $g_1(s,t)$; (ii) in general, the famous four slopes for $g_1(s,t)$ predicted by Doi and Edwards are preserved; however, depending on the value of the drift $c$, the escape of the chain from the tube can occur before $\tau_R$ and therefore some of the power laws predicted by the tube theory may disappear.

\begin{figure*}[hbt!]
	\centering
	\includegraphics[width=\columnwidth]{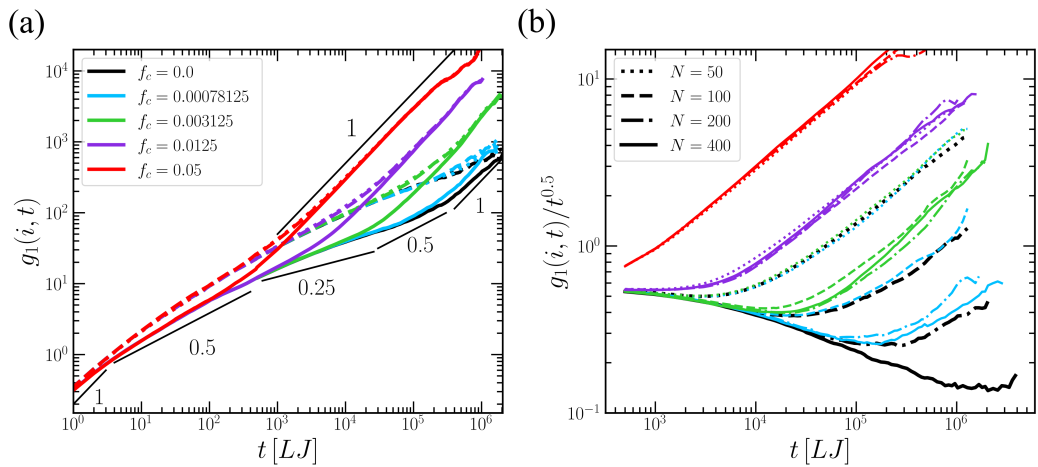}
	\caption{(a) Mean square displacement of the middle bead (solid lines), head (dashed lines), and tail (dotted lines) for different values of drift and polymer length $N=200$. Black solid lines highlight the power laws predicted by the reptation theory\cite{doi1988}. (b) MSD of the central monomer divided by $t^{0.5}$ for the same activities presented in panel (a) and different molecular weights: $N=50$ (dotted line), $N=100$ (dashed lines), $N=200$ (dotted-dashed lines), and $N=400$ (solid lines).}
	\label{fig: Figg1}
\end{figure*}

The theory and Brownian simulations consider the 1D dynamics of a chain (of constant and fluctuating length, respectively) moving along the primitive path, which is a 3D random walk. In molecular dynamics simulations, none of these assumptions is enforced, and therefore deviations from the predicted $g_1$(s,t) may occur. To explore this point, we first investigate the end monomers (head and tail) and middle monomers for different activities (see Fig. \ref{fig: Figg1}a)). As can be seen, the heads (dashed lines) and tails (dotted lines) of the polymer are hardly distinguishable for all the explored values of $f_c$, in agreement with both the theory and the BD simulations. Thus, although in MD the monomers have some lateral fluctuations inside the tube, the head-tail symmetry predicted by the theory with respect to the segmental motion is preserved. 

Furthermore, the four classical slopes $t^{0.5}$ (before $\tau_e$), $t^{0.25}$ (between $\tau_e$ and $\tau_R$), $t^{0.5}$ (between $\tau_R$ and $\tau_d$) and $t^1$ (beyond $\tau_d$) are clearly visible for the passive case (as expected and already reported many times in the literature \cite{Kremer_1990,P_tz_2000,likhtman_2007}). However, as the activity increases progressively, some of the intermediate power laws are lost. For example, when the activity induces a terminal time that is smaller than the Rouse time (green curve), the final slope $t^{0.5}$ completely vanishes and the middle monomer crosses over from $t^{0.25}$ directly to $t^1$. In addition, for very fast drifts (red and purple curves), the terminal time gets close to $\tau_e$ and the $t^{0.25}$ regime is almost completely lost. Again, even though the chain length is moderate ($N=200$, which corresponds to approximately 3 entanglements) and that in molecular dynamics simulations there is much more freedom for monomers and entanglements to fluctuate, the predictions of the theory regarding the form of $g_1$ are fully reproduced.

In Fig. \ref{fig: Figg1}b), the MSD of the central monomer $g_1(1/2,t)$ of chains with molecular weights ranging from 50 to 400 are shown, for the same activities as in panel (a). As expected, for passive chains, the monomeric MSD slows down considerably as the molecular weight increases. However, as the activity grows, this difference due to the molecular weight becomes narrower, to the point that the segmental MSD is almost indistinguishable for chains with different lengths and large activities. Unexpectedly, the polar activity introduced in this work not only makes the diffusion coefficient (an averaged quantity related to the motion of the center of mass) independent of the molecular weight, but also the segmental motion at intermediate times (a much more detailed quantity) is indistinguishable for chains of different molecular weights.

\subsection{End-to-end relaxation}\label{subsec: EteRelax}

\begin{figure*}[hbt!]
	\centering
	\includegraphics[width=\columnwidth]{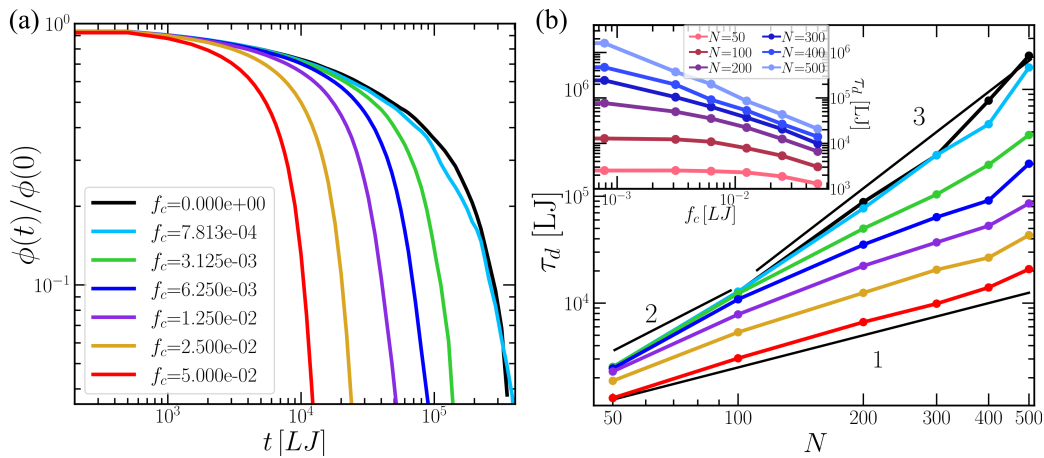}
	\caption{(a) End-to-end relaxation of active chains with $N=200$ for different values of the activity. Note that $\phi(t)$ is normalized by its value in equilibrium $\phi(0)$, although the mean end-to-end distance is independent of the activity in this work. (b) Relaxation time of all systems as a function of the molecular weight $N$, for the same set of drifts shown in panel (a), and as a function of the activity (inset).}
	\label{fig: Figend}
\end{figure*}

The end-to-end relaxation is often used to characterize the dynamics of entangled polymers because it is one of the slowest relaxation quantities in the system and because it can be related to experimental measurements of dielectric spectroscopy \cite{Watanabe_1994, Watanabe_2005a}. The end-to-end autocorrelation function can be defined as:
\begin{equation}
	\phi(t) = \langle \vec{R}(t') \cdot \vec{R}(t'+t) \rangle,
\end{equation}
where $\vec{R}$ is the end-to-end vector and the average is taken over all chains in the system and all possible time origins $t'$. In Fig. \ref{fig: Figend}a), the end-to-end relaxation of active chains with $N=200$ and different values of the drift is shown. As expected, the end-to-end relaxation is slower for passive chains (black curve) and shows a characteristic almost single-exponential decay at the disengagement time $\tau$. As the activity increases, relaxation occurs earlier, and the terminal decay becomes sharper. We can fit the data to a single exponential to extract the terminal time $\tau$ of each curve (details are provided in section SIII of the Supplementary Material, and the values of all parameters are provided in Table S1). 

If we plot the terminal time $\tau_d$ as a function of the molecular weight $N$ we expect to obtain the typical Rouse scalings ($\tau_d\propto N^2$), when the system is unentangled ($N\lesssim100$) and the reptation dependence ($\tau_d\propto N^{3.4}$) for entangled polymers ($N\gtrsim 200$). However, as the activity increases, the terminal time slowly departs from the classical polymeric behavior, becoming proportional to the molecular weight $N$, to the point that, for very relevant drifts ($\phi=0.05$), the terminal time grows linearly with $N$, as predicted by the theory \cite{Tejedor_2020,Tejedor2019}. More importantly, no signs of the unentangled to entangled transition can be observed. The linearity of $\tau_d$ with respect to $N$ is expected since, for a constant drift velocity, the time it takes for the chain to fully escape the tube is proportional to the length of the tube, which is proportional to the molecular weight. In the inset, the terminal time is plotted as a function of the activity $f_c$, where it can be seen that for short chains and very small activities, $\tau_d$ is almost independent of the activity. This is a sign that the activity is not sufficiently large to dominate over the reptation dynamics. On the other hand, as the molecular weight increases, even small values of the activity become relevant, and the terminal time starts to decay with the inverse of the activity, as expected. 

\subsection{Tangent-tangent correlation and tube survival}\label{subsec: TanTan}

\begin{figure}[hbt!]
	\centering
	\includegraphics[width=0.9\columnwidth]{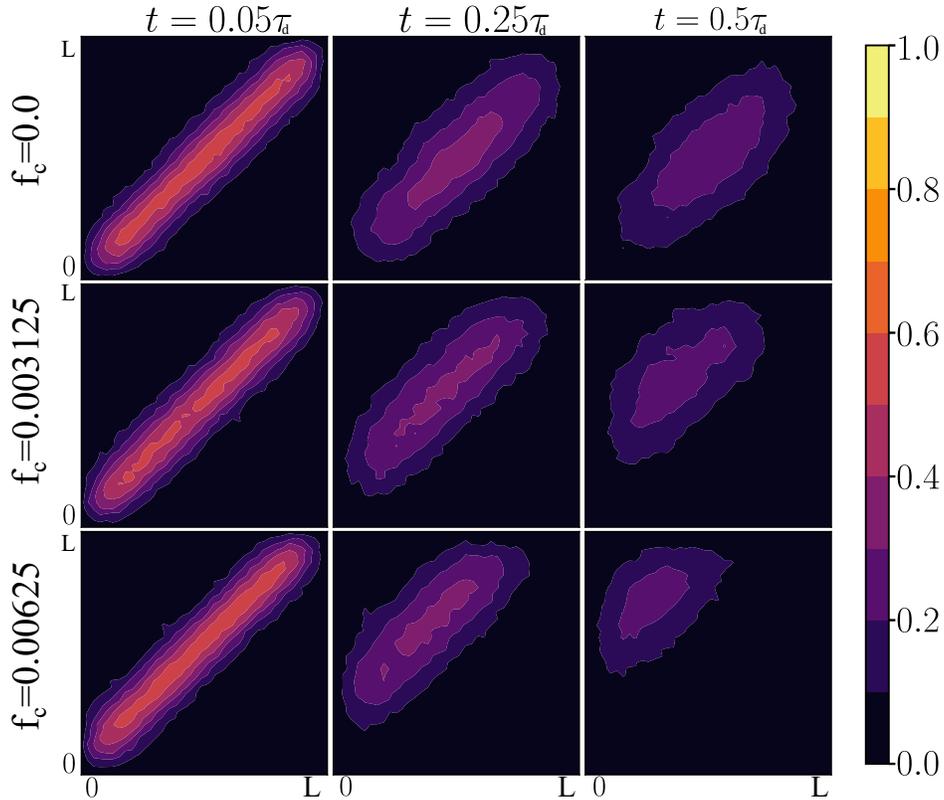}
	\caption{Tangent-tangent correlation function for chains of $N=200$ at different drift values ($f_c=0, 0.003125$ and $0.00625$, from top to bottom), evaluated at different times $t=0.05\tau_d,\, 0.25\tau_d$ and $0.5\tau_d$ (from left to right), where $\tau$ is the disengagement time of passive chains.}
	\label{fig: TanTan}
\end{figure}

Coarse-grained formulations of the tube theory \cite{Milner_2001, Graham_2003} strongly rely on the tube tangent correlation function, which relates the orientation of the vectors tangent at different points of the primitive path and at different moments in time, and can be used to calculate other observables such as the stress tensor, the single-chain structure factor, and the tube survival function\cite{doi1988,Tejedor_2020}. In MD simulations, the location of the primitive path is not readily available, and the tangent vectors of the atomistic chain fluctuate strongly. To overcome this problem, we have defined a coarse-grained version of the tangent-tangent correlation function, which is defined as:
\begin{equation}\label{eq: TanTanDefinition}
    G(i,i',t)=\langle \mathbf{u}(i,t)\cdot \mathbf{u} (i',0) \rangle,
\end{equation}
where the coarse-grained tangent vectors $\mathbf{u}(i,t)$ are calculated by considering monomers that are spaced by 4 bonds, to alleviate to some extent the fast decorrelation of the bond vectors due to CLF and lateral fluctuations inside the wide tube:
\begin{equation}\label{eq: TanVectorDefinition}
    \mathbf{u}(i,t)=\mathbf{r}_{i+5}(t)-\mathbf{r}_{i}(t),
\end{equation}
and the average is taken over all the active chains in the system and all possible origins of time. In Fig. \ref{fig: TanTan}, the tangent-tangent correlation function is shown for chains of $N=200$, different drift values, and different times. As expected, the tangent vector correlation function of passive chains (see Fig. \ref{fig: TanTan}, top row) remains symmetric with respect to the two diagonals at all times, because the head-tail symmetry is preserved by pure reptation, and the correlation slowly decays to zero as the correlation time approaches the terminal time. In contrast, the polar activity introduces a head-tail asymmetry that breaks the symmetry along the main diagonal of the tangent-tangent graph. This asymmetry is more pronounced for higher drift values and the correlation function decays faster. Still, the tangent correlation function remains symmetric with respect to the other diagonal. At a very early time ($t=0.05\tau_d$), the diffusive character of reptation dominates over the drift imposed by the active force, and the tangent correlation function is very similar to the passive case. As time progresses, the reptation dynamics is gradually replaced by the drift, and the correlation function decays faster. At $t=0.5\tau_d$, the tangent correlation function has almost decayed to zero in the fastest case shown in the figure. These results from Molecular Dynamics simulations qualitatively support all the predictions of the theory (see Fig. 9 in \cite{Tejedor_2020}). However, the coarse-grained vectors $\mathbf{u}(i,t)$, used to calculate $G(i,'i,t)$, can fluctuate within the tube and, thus, the decay at short times is more pronounced in MD simulations than in the theory. This effect is more clear in the tube segment survival function, which we calculate below.

\begin{figure}[hbt!]
	\centering
	\includegraphics[width=\columnwidth]{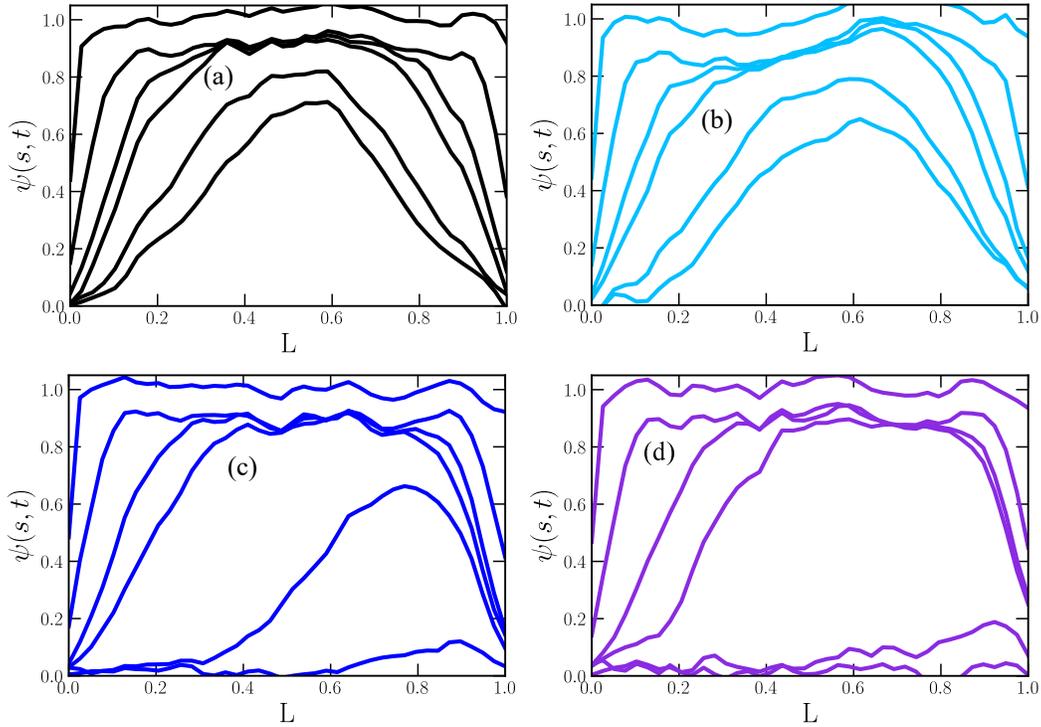}
	\caption{Tube segment survival function for chains with $N=200$ at drift values $f_c=0$ (a, black), $f_c=0.00078125$ (b, cyan), $f_c=0.00625$ (c, blue), and $f_c=0.0125$ (d, purple). The functions are evaluated at times $t=0,\,0.01,\,0.05,\,0.10,\, 0.5,\,1\times\tau_d$, where $\tau_d$ is the disengagement time in the passive case.}
	\label{fig: Surfun}
\end{figure}

By integrating the tangent correlation function $G(i,i',t)$ with respect to either of the two indices related to the position along the chain, another key quantity in the tube theory can be obtained: the tube segment survival function $\psi(s,t)$. This function represents the probability that a segment $s$ has not been visited by either of the two ends of the chain at time $t$. According to the theory, the drift should introduce a very clear asymmetry in $\psi(s,t)$, with tube segments being renewed faster by the chain tail than by the head, and a much faster decay to zero (see Figure 12 in \cite{Tejedor_2020}). These predictions are supported by the results of $\psi(s,t)$ from MD, shown at different times in Fig. \ref{fig: Surfun} for $N=200$ and different values of the activity. First, it is important to note that the curves of $\psi(s,t)$ are rather noisy compared to the theory, due to the small size of the ensemble of active chain used in each simulation. Second, the value of $\psi(s,t)$ decays clearly and homogeneously along the tube contour at very early time. This is due to the lateral freedom of the chains, which allows them to explore the whole width of the fat tube, whereas in the theoretical description the tube is infinitely thin and there is no such decay. Third, due to the finite size of the coarse-grained tangent vectors defined in this work, the segments closer to the ends are not immediately destroyed when they are visited by the ends, they take a finite time to reorient and lose the memory of their previous orientation. Thus, $\psi(s,t)$ for the end segments does not immediately decay to zero at times $t>0$. Regardless of these differences, the major trends predicted by the theory are captured by the MD simulations. The drift introduces a clear asymmetry in $\psi(s,t)$, with segments closer to the tail being more easily destroyed than those closer to the head, and the decay of $\psi(s,t)$ is faster than in the passive case, and increases for larger drifts. At $t=0.5\tau$, the entire tube has been almost completely renewed for the fastest drift shown in the figure. 

To end this section, it is important to mention that the integral of $\psi(s,t)$ with respect to $s$ yields the tube survival function $\psi(t)$, which provides the average fraction of the original tube that has not yet been visited by either of the ends of the chain at time $t$. In the tube theory, this function is equivalent to the relaxation of the end-to-end vector $\phi(t)$. 

\subsection{Dynamic structure factor}\label{subsec: DSF}

The dynamic structure factor can be used to study the relaxation dynamics of a material at different length-scales using scattering experiments \cite{leger1981reptation,likhtman2005,richter2005}. The dynamic structure factor can be calculated as \cite{doi1988}
\begin{equation}
    S(\mathbf{q},t)=\frac{1}{N}\sum_{n,m}\langle\exp[i\mathbf{q}\cdot (\mathbf{r}_n(t)-\mathbf{r}_n(0))]\rangle,
\end{equation}
where $\mathbf{r}_n$ is the position of the bead $n$, $\mathbf{q}$ is the scattering vector, and $i$ is the imaginary unit. The double sum in the latter equation runs over all the beads of the chain, and the average is taken over all chains and all possible time origins. In our case, we study the dynamic structure factor for $N=200$, since shorter chains are barely entangled and long distances are not sufficiently well explored for longer chains, resulting in poor statistics. In our previous work \cite{Tejedor_2020}, we discussed two different regimes depending on the magnitude of the drift velocity relative to that of the scattering vector. In the MD simulations, the exact value of the parameters that control the significance of $c$ and $\mathbf{q}$ can be expressed as
\begin{equation}
    \mu\sim\mathbf{q}^2N,\qquad \nu\sim \frac{cN}{D_G},
\end{equation}
where $D_G$ is the pure reptation diffusion coefficient. In our analysis, the scattering vectors (and corresponding characteristic distances) are chosen relative to the coil size as $\mathbf{q}^{-2}=d^{2}=0.01,0.032,0.1,0.32,1\times C_\infty\cdot N$, where $C_\infty=1.88$ is the characteristic ratio of the Kremer-Grest model\cite{Sliozberg_2012}. Smallest values of the scattering vector $\mathbf{q}^2$ (corresponding to longer distances $d^2\gtrsim 1$) cannot be properly measured because, in order to obtain good statistics, the simulation times are too long. 

\begin{figure}[hbt!]
	\centering
	\includegraphics[width=\columnwidth]{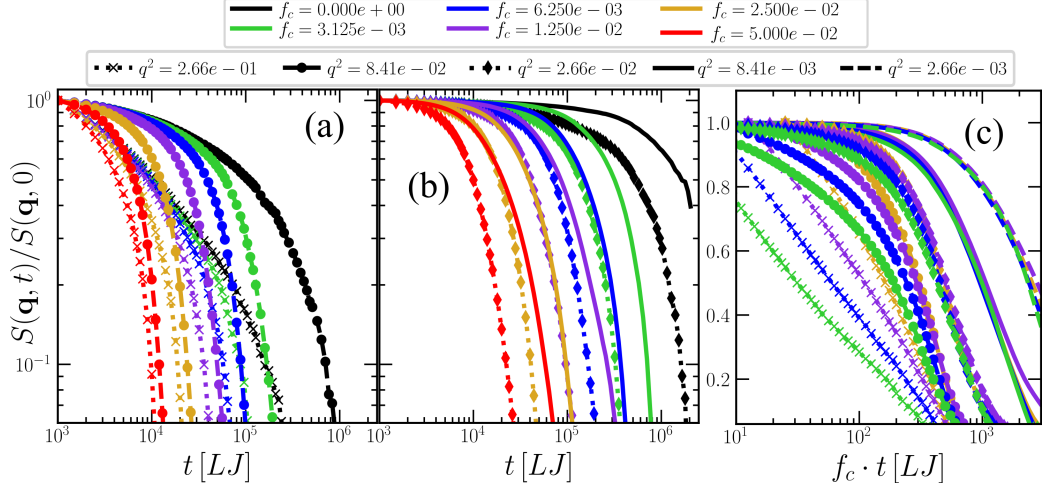}
	\caption{Dynamic structure factor of active chains with $N=200$ for different activities and scattering vectors that explore short distances (Panel a) and intermediate long distances (Panel b). (c) Dynamic structure factor for selected drift values vs $f_c\cdot t$ to show the limit $\nu\gg\mu$}
	\label{fig: DSF}
\end{figure}

The theory predicts two different regimes. First, when the drift is significant and we explore short distances, the dynamic structure function is proportional to the tube survival factor is proportional to the tube survival function or the end-to-end relaxation, i.e., $S(\mathbf{q},t)\propto\psi(t) \propto \phi(t)$ (see Eq. (35) in the theoretical work~\cite{Tejedor_2020}). 
As can be seen in Fig. \ref{fig: DSF}a), the dynamic structure factor of a chain with $N=200$ exhibits faster relaxation at short times than the end-to-end vector in Fig. \ref{fig: Figend}. This is to be expected, because the width of the virtual tube in MD simulations, compared with the infinitely thin tube of the theory, allows the chain segments to faster relax at short distances within an entanglement. In addition, it can be seen that the effect of the activity at short distances , ($\mathbf{q}^2\sim0.1$, see Figure \ref{fig: DSF}a) is more moderate than at longer distances, which confirms that short distances are dominated by pure reptation. 

The second regime predicted by the theory is given in the limit of relevant drifts and distances that can be short or long but must meet the condition $\nu\gg\mu$. In this limit, the dynamic structure function decays as a single-exponential (see Eq. (38) in~\cite{Tejedor_2020})
\begin{equation}
    S(\mathbf{q},t)\propto \exp\left(-\frac{\mathbf{q}^2ct}{6}\right),
\end{equation}
with a relaxation time that is inversely proportional to drift or the activity ($c\propto f_c$). In Fig. \ref{fig: DSF}(b), we show the results for this regime, showing an almost single-exponential decay with terminal times that are equally spaced, indicating a relaxation time that is proportional to $f_c$. To better highlight this result, all the curves shown in Fig. \ref{fig: DSF}(a) and (b) are shown in Fig. \ref{fig: DSF}(c), but scaling the time axis by $f_c$, to prove that the terminal time is proportional to $f_c$ for small scattering vectors. It can be seen that curves of $S(\mathbf{q},t)$ do not overlap for large $\mathbf{q}$ vectors (when the condition $\nu\gg\mu$ does not apply), but the overlap gets increasingly better as $\mathbf{q}$ decreases, supporting the prediction from the theory. 

\section{\label{sec:conclusion}Summary and Conclusions}

In this work, we run molecular dynamics simulations of a modified Kremer-Grest model to explore the dynamics of entangled polymers subjected to a polar activity that acts tangentially to the chain contour and to verify the most important predictions of a previously published analytical theory of active entangled polymers. First, we prove that our choice of active force applied tangentially to the chain contour qualitatively has the same effect as the active drift imparted to the primitive path in the theory. Then, we show that, for entangled polymers, there is a wide range of activity values that strongly affect the dynamical behaviour of chains but preserve all the equilibrium static properties related to the chain and the tube. Finally, all the predictions of the theory are confirmed, as follows: i) the segmental mean-squared displacements of the head and tail monomers are identical; ii) the MSD of the central monomer shows the classic characteristic power laws of the tube theory, which progressively dissappear as the activity increases; iii) the MSD of the center of mass shows a transient super-diffusive scaling; iv) the self-diffusion coefficient becomes independent of the molecular weight and proportional to the activity, for relevant values of the drift; v) the end-to-end relaxation decays faster and with a sharper terminal region as the activity increases; vi) with a terminal time that is proportional to the molecular weight and inversely proportional to the activity; vii) the tangent-tangent correlation function and the tube segment survival function become highly asymmetrical and relax much faster with the activity; and viii) the dynamic structure factor shows very rich relaxation behaviour which is confirmed qualitatively by the MD simulations.  

The agreement of the theory predictions with the results from MD simulations is excellent considering that some of the theory assumptions are not respected in the simulations, i.e., the tube is not infinitely thin, and the ends of the chains do not forget their previous orientations immediately. Having an accurate theoretical framework to study the dynamics of entangled active polymers can be very useful for both experiments and theories of active polymers. Our results can be used to guide the study of active biomolecules or to help design new macromolecular materials with rich and enhanced dynamical properties.

\section*{Acknowledgements}

A.R.T. is funded by Universidad
Politécnica de Madrid (PhD fellowship ‘programa propio UPM’).
The authors acknowledge funding from the Spanish Ministry of Economy
and Competitivity (PID2019-105898GA-C22) and the Madrid
Government (Comunidad de Madrid-Spain) under the Multiannual
Agreementwith Universidad Politécnica deMadrid in the line Excellence
Programme for University Professors, in the context of the V PRICIT
(Regional Programme of Research and Technological Innovation).
The authors gratefully
acknowledge Universidad Politécnica de Madrid (\url{www.upm.es}) for providing computing resources on Magerit Supercomputer.



\bibliographystyle{elsarticle-num}
\bibliography{cas-refs}





\end{document}